\documentclass{epl}
\usepackage{array}
\usepackage{amsmath,amssymb,amsfonts}
\usepackage{graphicx}

\def\0{\varnothing}

\title{Dynamical origin of spontaneous symmetry breaking in a
field-driven nonequilibrium system}
\shorttitle{Spontaneous symmetry breaking}
\author{R. D. Willmann\inst{1} \and G. M. Sch\"utz\inst{1} \and S. Gro{\ss}kinsky\inst{2}}
\institute{
  \inst{1} Institut f\"{u}r Festk\"{o}rperforschung, Forschungszentrum
  J\"{u}lich, 52425 J\"{u}lich, Germany\\
  \inst{2} Zentrum Mathematik, TU M\"{u}nchen,
  85747 Garching bei M\"{u}nchen, Germany
}

\shortauthor{R. D. Willmann et al.}


\begin{document}

\maketitle

\thispagestyle{plain}

\begin{abstract}
A one-dimensional driven two-species model with parallel sublattice
update and open boundaries is considered. Although the microscopic
many-body dynamics is symmetric with respect to the two species and
interactions are short-ranged, there is a
region in parameter space with broken symmetry in the steady state.
The sublattice update is deterministic in the bulk and allows for a
detailed analysis of the relaxation dynamics, so that symmetry
breaking can be shown to be the result of an amplification mechanism
of fluctuations.
In contrast to previously considered models, this
leads to a proof for spontaneous symmetry breaking which is valid
throughout the whole region in parameter space with a symmetry broken steady state.
\end{abstract}

While single-species driven diffusive systems in one dimension are
largely understood, two-species models show a variety of phenomena
that are a matter of current research, such as phase separation and
spontaneous symmetry breaking (see \cite{Schutz03} for a recent review).
The first such model that was shown to exhibit spontaneous symmetry
breaking was a model with open boundaries that became
known as the 'bridge model' \cite{Evans95}. In this model, two species
of particles move in opposite directions. Although the dynamical rules
are symmetric with respect to the two species, two phases with
non-symmetrical steady states were found by Monte Carlo
simulations and mean-field calculations. While the existence of one of
the phases remains disputed \cite{Clincy01,Arndt98}, a proof for the
existence of the other one was given for the case of one vanishing
boundary rate \cite{Godreche95}.
Recently, a variant of the bridge model with non-conserving bulk dynamics
was considered \cite{Levine04}. Although the phase diagram of this model
is even richer than that of the original one, a proof for a symmetry
broken state could again only be given in the case of one vanishing
boundary rate. Since spontaneous symmetry breaking does not occur in
one-dimensional systems in thermal equilibrium and since not even a
macroscopic description in terms of boundary reservoirs is known
\cite{Popk05,Toth03}
it would be desirable to
gain insight in the dynamical origin of this genuinely nonequilibrium
phenomenon.

The symmetry breaking models considered so far evolve by random
sequential update. In this article, a variation of the bridge model
with parallel sublattice update is studied. The update scheme
ensures that the dynamics in the bulk is deterministic, while
stochastic events occur at the boundaries. Thus -- while maintaining
noisy dynamics -- the complexity of the
problem is reduced, which allows to elucidate the mechanism by which
spontaneous symmetry breaking occurs in this model as well as to give
a proof for the existence of a symmetry broken phase. This proof is
valid for the whole region in parameter space where symmetry breaking
occurs and not just in some limiting case. 

The model considered here is defined on a one-dimensional lattice of
length $L$, where $L$ is an even number. Sites are either empty or
occupied by a single particle of either species $A$ or $B$, i.e.,
the particles are subject to an exclusion interaction.
The dynamics is defined as a parallel sublattice update scheme
in two half steps.
In the first half-step the following processes take place: At site $1$
it is attempted to create a particle of species $A$ with
probability $\alpha$ if the site is empty, or to annihilate a particle
of species $B$ with probability $\beta$, provided the site is occupied by
such a particle:
\begin{equation}
0 \overset{\alpha}{\to} A \quad B \overset{\beta}{\to} 0\ .
\end{equation}
At site $L$, a particle of species $B$ is created with probability
$\alpha$ and a particle of species $A$ is annihilated with probability
$\beta$:
\begin{equation}
0 \overset{\alpha}{\to} B \quad A \overset{\beta}{\to} 0\ .
\end{equation}
In the bulk, the following hopping processes occur deterministically
between sites $2i$ and $2i+1$ with $0<i<L/2$:
\begin{equation}
A0 {\to} 0A \quad 0B {\to} B0 \quad AB {\to} BA.
\end{equation}
In the second half-step, these deterministic hopping processes take
place between sites $2i-1$ and $2i$ with $0 < i \leq L/2$. Note that
the dynamics is symmetric with respect to the two particles species.
The original bridge model \cite{Evans95} arises as the continuous-time
limit of this model with stochastic hopping.

\begin{figure}
\begin{center}
  \includegraphics[width=6.6cm]{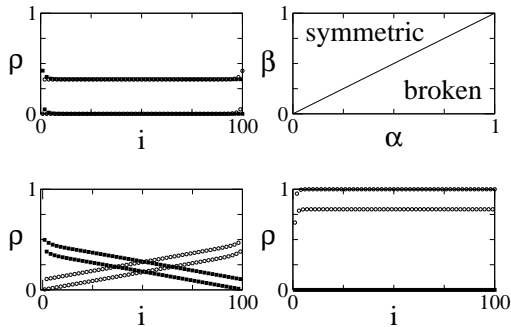}
  \caption{Upper right: Stationary phase diagram of the sublattice bridge
    model. Average density profiles as obtained from Monte Carlo
    simulations in the symmetric phase at $\alpha=0.6$ and $\beta=0.8$
    (upper left), the broken phase at $\alpha=0.6$ and $\beta=0.2$
    (lower right) and on the transition line at $\alpha=\beta=0.2$
    (lower left). $A$ densities are shown by $\circ$ and $B$ densities
    by $\scriptstyle\blacksquare$.}
\label{fig:phasediagram}
\end{center}
\end{figure}

The stationary phase diagram of the model in terms of the parameters
$\alpha$ and $\beta$ can be explored by Monte Carlo
simulations. Two phases are found (see Figure \ref{fig:phasediagram}):
\begin{itemize}
\item If $\alpha<\beta$, the system exhibits a symmetric steady
state. Here, the bulk densities are $\rho_A(i)=0$, $\rho_B(i)=\alpha
\beta /(\alpha+\beta)$ if $i$ is odd, and
$\rho_A(i)=\alpha \beta /(\alpha+\beta)$, $\rho_B=0$ if $i$ is even.
\item If $\alpha>\beta$, the system resides in the symmetry broken
  phase. Assume the $A$ particles to be in the majority. Then, the
  bulk densities in the steady state are $\rho_B(i)=0$ for all $i$,
  $\rho_A(i)=1$ for $i$ even and $\rho_A(i)=1-\beta$ for $i$ odd. This
  means that the symmetry is maximally broken and the minority
  species is completely expelled from the system.
\item The behavior on the transition line for $\alpha =\beta$ is
  described below.
\end{itemize}
Thus, the dynamics of the majority species in the broken phase is identical
  to the single species ASEP with parallel sublattice update. For this
  system, the exact steady state density profile in a finite system is known
  \cite{Schutz93}. The density profile of the majority species in the
  broken phase of the sublattice bridge model equals that of the
  high density phase in the sublattice ASEP at the given parameters
  $\alpha$ and $\beta$.

In the following, the dynamics leading to symmetry breaking is
elucidated and the respective time scales are determined. For $\alpha
>\beta$ an amplification mechanism of fluctuations leading to a
symmetry broken state is identified. This mechanism
shows that the system needs a time $T_1$ which is only
algebraically increasing with $L$ to enter the broken phase, if it was
started with symmetric initial conditions.
Furthermore, assuming the system to be in the broken phase it is shown
that it takes a time $T_2$ that is exponentially increasing with $L$
until particles of the minority species can penetrate the system.
Both facts together provide a proof for spontaneous symmetry breaking
in this model for $\alpha >\beta$.

{\it Dynamics of symmetry breaking: }
It is assumed that at $t=0$ there are no particles in the system and
that $\alpha>\beta >0$. The case of other initial densities can be
treated in a similar fashion and the case $\beta =0$ leads to a
degenerate situation with many steady states depending on the initial
conditions, since no particle can leave the system. This is
straightforward to analyze and is not discussed in the
following. Starting from the empty lattice, $A$ $(B)$ particles are created
at every time step with probability $\alpha$ at site $1$ $(L)$. Once
injected, particles move deterministically with velocity $2$
$(-2)$. Therefore, at time $t=L/2$ the system is in a state where
the density of $A$ $(B)$ particles is $\alpha$ $(0)$ at all even sites and $0$
$(\alpha)$ at all odd sites.
In this situation both creation and annihilation of particles are
possible.

However, it turns out that the effect of creation of particles
is negligible:
Since $\alpha >\beta$ the deterministic hopping with velocity $2$
transports on average more $A$-particles
towards site $L$ than can be annihilated there. This leads to the formation of
an $A$-particle \textbf{jam} at the right boundary, blocking the injection of
$B$-particles. An analogous argument holds for the left boundary,
which is blocked by a $B$-particle jam. In these jams, the only source
of vacancies is annihilation at the boundaries with probability
$\beta$ in the first half-step. In the second half-step the vacancy
moves one site towards the bulk with probability $1$. Therefore, in a jam, the
density of $A$ $(B)$ particles at even (odd) sites is 1, while that at
odd (even) sites is $1-\beta$. So the only way to create particles in
this situation is a complete dissolution of a jam. But as long as it
gains particles from the low density region this is a very rare event
since $\alpha >\beta$. It can be shown that the average number of
created particles is small and bounded independent of $L$ \cite{todo}.
So creation of particles in this jammed situation becomes negligible
in the limit $L\to\infty$ and will be neglected in the following.

\begin{figure}
\begin{center}
\includegraphics[width=4.5cm,angle=0]{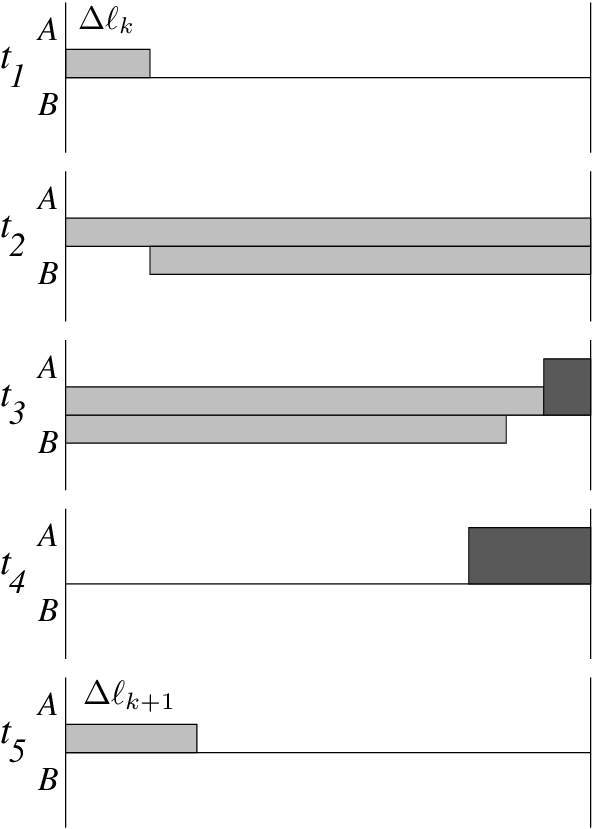}\hspace{10mm}
\includegraphics[width=8cm,angle=0]{SSB2.eps}
\caption{Left: Illustration of the stages involved in the $k$-th cycle of
    the dynamics of spontaneous symmetry breaking as explained in the
    text. Low density regions are drawn light grey, jams are
    dark grey and white regions of the system are empty.\newline
Right: Symmetry breaking, starting from the empty lattice
(single realization of MC simulation). Here, $\alpha=0.9$, $\beta=0.8$ and $L=10000$. The density of $A$ particles is drawn in black, that of $B$ particles in grey and the difference as the thick black line. The inset shows
a close-up during the time evolution. Individual stages as described
in the text are separated by dotted vertical lines.
}
\label{fig:stages}
\end{center}
\end{figure}

The number of particles in each of the two jams reduces by one in
every time step with probability $\beta$. Since creation of
particles is negligible, the influx into the jam
ceases after some time and the jam eventually dissolves. By
fluctuations, one of the jams, say the $B$-jam at the left boundary,
dissolves first. $A$ particles can enter the system while $B$
particles are still blocked until the $A$-jam at the right boundary is
also dissolved.
The configuration of the system at this point is illustrated in Figure
\ref{fig:stages} ($t_1$). The light grey region denotes a
\textbf{region of low density} of $A$ particles where the density is
$\alpha$ $(0)$ on even (odd) sites. The (random) length of this
region, $\Delta\ell_1$, describes the majority of one of the
species. The description is symmetric, so if $\Delta\ell_1 <0$ this
corresponds to a majority of $B$ particles.
Thus the average value $\langle\Delta\ell_1 \rangle =0$, but typically
$\Delta\ell_1 =O(\sqrt{L})$ due to Gaussian fluctuations for large $L$
and one of the species has the majority, which we assume to be species
$A$.

The time evolution just described
constitutes the first loop of a cyclic behaviour which can be effectively
described by the dynamics of low density regions and jams at the
boundaries. The key ingredient for this simplification is the jamming
mechanism described above. The cyclic behaviour consists of 4 stages,
which we summarize in the following and which is illustrated
in Figure \ref{fig:stages}.

\begin{enumerate}
\item At the beginning of a cycle ($t_1 =0$) there is a low density
  region of $A$ particles at the left boundary of length
  $\Delta\ell_k$. Both species enter the system with probability
  $\alpha$ and penetrate the bulk deterministically with speed $2$ $(-2)$.
\item The region of $A$ particles reaches the right boundary at
  time $t_2 =\big( L-\Delta\ell_k \big) /2$, creation of $B$ particles is
  blocked. $A$ particles still enter with probability $\alpha$ and
  exit with probability $\beta <\alpha$, further increasing their
  majority.
\item At time $t_3 =L/2$ the $B$ particles reach the left boundary,
  blocking also the creation of $A$ particles. Both species form jams
  at the boundaries, which gain particles from the low density regions.
  Since creation of particles at the boundaries is negligible, both
  jams eventually dissolve.
\item Let $t_4 \geq t_3$ be the time when the jam of $B$ particles is dissolved
  and $A$ particles start to enter the system. Again, since $\alpha
  >\beta$ the majority of $A$ particles increases on average.
\item At time $t_5 \geq t_3$
  the $A$-jam at the right boundary is dissolved and also $B$
  particles can enter the system.
\end{enumerate}
The cycle is finished when both jams are dissolved. In Figure
\ref{fig:stages} it is assumed that the $B$-jam dissolves first,
i.e.\ $t_4 <t_5$, which is most likely if $\Delta\ell_k >0$. But in
general $t_4 \geq t_5$ is also possible and included in the above
description. The result of the cycle is
\begin{equation}\label{result}
\Delta\ell_{k+1} =2\, (t_5 -t_4 )\ ,
\end{equation}
which is the initial condition for the next loop. If $\Delta\ell_{k+1}
<0$, $A$ and $B$ particles have to be interchanged for the next cycle
in the above description. Note that within this framework the
process starting from the empty lattice is a cycle with
initial condition $\Delta\ell_0 =0$ and $t_2 =t_3$.

In the following we analyze the distributions of the random
variables $t_4$ and $t_5$ to get the time evolution of
$\Delta\ell_k$. Let $\tau_n$ be the (random) time it takes for a jam
of length $n$ to dissolve. With this
\begin{equation}\label{t4t5}
t_4 =L/2+\tau_{N_B}\ ,\qquad t_5 =\big( L -\Delta\ell_k \big) /2 +\tau_{N_A}\ ,
\end{equation}
where $N_A$ ($N_B$) denotes the number of
$A$ ($B$) particles that entered the system up to time $t_3$ ($t_2$)
before blocking, including $\Delta\ell_k$. 
These are Bernoulli random variables with hitting probability $\alpha$,
and thus the average values are
\begin{eqnarray}\label{nanb}
\langle N_A \rangle &=&\big(\Delta\ell_k /2+t_3 \big)\,\alpha
=\big(L+\Delta\ell_k \big)\,\alpha /2\ ,\nonumber\\
\langle N_B \rangle &=& t_2 \,\alpha =\big(L-\Delta\ell_k
\big)\,\alpha /2\ .
\end{eqnarray}
As the boundary site in a jam is always occupied and particles are
annihilated with probability $\beta$, the time
$\tau \in\{1,2,\ldots \}$ for a particle to leave the jam is a
geometric random variable with parameter $\beta$. It is
$\langle\tau\rangle =1/\beta$ and thus the average value for $n$
particles to leave is $\langle\tau_n \rangle =n/\beta$.
So dividing (\ref{nanb}) by
$\beta$ yields $\langle\tau_{N_A}\rangle$ and
$\langle\tau_{N_B}\rangle$. Using this and (\ref{result}) to (\ref{nanb})
the average value of $\Delta\ell_{k+1}$ conditioned on $\Delta\ell_k$
can be computed, and
\begin{equation}\label{drift}
\big\langle \Delta\ell_{k+1} -\Delta\ell_k \,\big|\,\Delta\ell_k
\big\rangle= 2\,\Big(\,\frac\alpha\beta -1\Big)\,\Delta\ell_k +O(1)\ .
\end{equation}
Here the correction terms due to rare fluctuations in the boundary
jams are included. They are shown to be $O(1)$ in a rigorous
treatment \cite{todo},
whereas $\Delta\ell_k$ is typically of order
$\sqrt{L}$ or larger. With (\ref{drift}), conditioned on the initial
fluctuation of $\Delta\ell_1$, the average value after $k$ loops is
\begin{equation}\label{amp}
\big\langle\Delta\ell_{k+1} \,\big|\,\Delta\ell_1 \big\rangle=
\big(\Delta\ell_1 +O(1)\big)\,\Big( 2\,\frac\alpha\beta -1\Big)^k\ .
\end{equation}
As $q=\big( 2\,\frac\alpha\beta -1\big) >1$ this constitutes an
amplification of the initial fluctuations $\Delta\ell_1 =O(\sqrt{L})$.

$\Delta\ell_k$ can be
interpreted as a random walker on $\{ -L,\ldots ,L\}$. The jump
probabilities are in principle given with (\ref{result}) and
(\ref{t4t5}) but such a detailed (and cumbersome) analysis is not necessary.
Since $\alpha >\beta$ the walker is driven towards the boundaries with drift
(\ref{drift}) proportional to $\Delta\ell_k$. As long as $-L<\Delta\ell_{k+1}
<L$ the cycle can restart with $0<t_2 \leq t_3$, and all the stages
are well defined within a rigorous treatment \cite{todo}.
Thus initial fluctuations of $\Delta\ell_1 =O(\sqrt{L})$ are
amplified, and when $\big|\Delta\ell_k \big|$ reaches system size
the amplification loop
stops.
In this situation, the effect
of injection of $B$-particles is negligible for large $L$ due to
jamming of $A$-particles at the right boundary. The only relevant
stochastic boundary processes are injection of $A$-particles at site
$1$ and annihilation of $A$ at site $L$. As there is a constant
supply of $A$ particles the jam quickly fills the system within a time
of order $L$, reaching a state as for the single-species ASEP
\cite{Schutz93} with sublattice update at the given $\alpha$, $\beta$.

{\it Time to reach the broken state: }With (\ref{amp}) it is
\begin{equation}\label{number}
\big\langle\Delta\ell_{k+1} \,\big|\,\Delta\ell_1 \big\rangle\geq
L\quad\mbox{if}\quad k\geq \frac{\ln L-\ln \big(\Delta\ell_1
  +O(1)\big)}{\ln q} =O(\ln L)\ .
\end{equation}
and we denote by $\bar{k}$ the number of cycles until (\ref{number})
for the average value is fulfilled. But in general the number of
cycles until the amplification loop ends is a random variable.
Analogous to (\ref{drift}) one can get a recursion relation for the
standard deviation $\sigma \big(\Delta\ell_k \big)$ \cite{todo} and
show that $\sigma \big(\Delta\ell_{\bar{k}} \big) =o(L)$. Thus
fluctuation are small compared to the average value, and for
$L\to\infty$ one has the expansion
\begin{equation}\label{expansion}
\frac{\Delta\ell_{\bar{k}}}{L} =
\frac{\big\langle\Delta\ell_{\bar{k}}\big\rangle}{L} +
\xi\frac{\sigma \big(\Delta\ell_{\bar{k}} \big)}{L} =1+\xi\, o(1)\ ,
\end{equation}
for almost every realization of the process, where $\xi$ is a random
variable of order $1$. In turn the average number of cycles to reach
the end of the amplification loop is $\bar{k}$ as determined by
condition (\ref{number}), and moreover, fluctuations vanish in the
limit $L\to\infty$.\\
The average length of a loop is given by
\begin{equation}\label{looplength}
\langle t_5 \rangle =\Big(\frac\alpha\beta +1\Big)\frac L2
+\Big(\frac\alpha\beta -1\Big)\, \big\langle\Delta\ell_k
\,\big|\,\Delta\ell_1 \big\rangle /2 =O(L)\ .
\end{equation}
The second term depending on the index $k$ is negligible unless for
$k$ close to $\bar{k}$, and even then the prefactor is smaller then
the one of the first term. Moreover, since $t_5$ is given by a sum of
independent random variables, fluctuations are Gaussian of order
$\sqrt{L}$ for large $L$, and thus the length of a loop is almost
constant, independent of $k$. This is confirmed in Figure \ref{fig:stages}
where a single realization of the process is plotted.
Since (\ref{expansion}) is also conditioned on the
initial fluctuation $\Delta\ell_1$, fluctuations of this quantity
constitute the only source of randomness in the time evolution for
large $L$.

In total, the average time to reach the symmetry broken
phase is typically $T_1 =O(L\ln L)$.
Thus, by {\it amplification of fluctuations the system reaches
a state of broken symmetry in a time algebraically increasing with $L$,
provided that $\alpha>\beta$}.
The broken states that are attained are the steady states of the
single species ASEP at the respective $\alpha$, $\beta$. These
states either contain only $A$- or $B$-particles, depending on
initial fluctuations.

{\it Residence time in the broken state: }Assume the system to be
residing in the broken state with particle species $A$ in the
majority. For $L\to\infty$ this means $\rho_A(i)=1+o(1)$ for even
sites and $\rho_A(i)=1-\beta +o(1)$ for odd sites, up to boundary
effects at the left boundary with $i=O(1)$. Species $B$ is expelled
from the system and
injection of $B$ particles is only possible if site $L$ is empty.
Exact expressions given in \cite{Schutz93} (equation (18)) yield
\begin{equation}
\rho_A (L)=1-\Big( 1-\frac{\beta\, (1-\alpha)}{\alpha\,
  (1-\beta)}\Big)\, \Big(\frac\beta\alpha\Big)^L \ .
\end{equation}
Thus the probability that site $L$ is empty is exponentially small
in the system size, and {\it the time $T_2$ until the minority
species can penetrate a system started in the broken state is
exponentially large in $L$.} This is not surprising even without
knowledge of the exact expressions, since for injection of the first
$B$ particle the complete jam of $A$ particles has to be dissolved
against the drive $\alpha >\beta$. This jam consists of the order of
$L$ particles. Together with the statement about $T_1$ from above,
this proves spontaneous symmetry breaking.

The amplification mechanism outlined above does not apply
for $\alpha <\beta$ since the formation of boundary jams, a key ingredient
for the amplification mechanism, does not work. The length of a
boundary jam is driven towards small values so the boundary sites are
not blocked and particles are injected all the time.

{\it Dynamics on the transition line: }For the borderline case
$\alpha =\beta$ the end of a boundary jam is diffusing and can take
very large values. So the cyclic behaviour can still be observed,
but fluctuations are larger and the cycle lengths, though still of
order $L$, are strongly fluctuating. But even if this effective
description is still valid, according to (\ref{drift}) there is no
amplification of fluctuations during a cycle. Instead,
$\Delta\ell_k$ is not driven towards the boundary but is diffusing,
so a symmetry broken state can still be reached within $O(L^2)$
cycles, and thus $T_1 =O(L^3)$. On the other hand, when the system
is in one of the symmetry broken states, the length of the jam of
the majority species is only diffusing. So it dissolves in a time of
only $T_2 =O(L^2)$, which is the lifetime of a symmetry broken state
for $\alpha =\beta$. Thus, no symmetry breaking takes place in this
case. Instead, for large $L$ a typical configuration is taken from a
cycle, consisting of jams with diffusing length and of low density
regions for both species. An average over many realizations leads to
an approximately linear density profile as shown in Figure
\ref{fig:phasediagram} (lower left). Further, for $\alpha =\beta$
site $2$ ($L$) is occupied by $A$ particles for approximately half
of a cycle length with probability $\alpha$ ($1$), leading to
$\rho_A (2)=\alpha /2$ and $\rho_A (L) =1/2$. For odd sides an
analogous argument yields $\rho_A (1)=0$ and $\rho_A (L{-}1)
=(1{-}\alpha )/2$, which agrees well with Figure
\ref{fig:phasediagram}. Moreover, the formation and dissolution of
boundary jams for the two species shows interesting temporal
correlations \cite{Popk04}.

It is also very interesting to compare these findings to the results of a
mean-field treatment of the system. Performing a similar analysis as
in \cite{Evans95} yields the same results for the steady state as the
above analysis, except for the transition line $\alpha =\beta$.
Along this line, mean field theory predicts a second symmetry-broken
phase, where the density of $A$ ($B$) particles in the bulk is
$\alpha_1$ ($0$) on even sites and $0$ ($\alpha_2$) on odd sites, with
the constraint $\alpha_1+\alpha_2=\alpha$.
This prediction corresponds to the disputed
low-asymmetric phase of the original bridge model
\cite{Evans95,Clincy01,Arndt98}, where for the present model only the sum
of the $A$- and $B$-densities is fixed.
In the present model, this phase does not exist as explained
above and the mean-field prediction $\alpha_1+\alpha_2=\alpha$ is in clear
contradiction to the observed density profile in Figure
\ref{fig:phasediagram} (lower left).

In the present article, a two-species driven model with deterministic
bulk behavior was investigated. The mechanism for spontaneous symmetry
breaking in this model was described, leading to estimates for the
relevant time scales in the broken phase  and proving the existence of
spontaneous symmetry-breaking without further assumptions  on the rates.
Along the transition line, the failure of a mean-field treatment,
leading to the prediction of a second asymmetric phase, was explicitly
demonstrated. In fact, the time evolution of the total density can be
described for both the symmetric and the asymmetric phase and will be
treated in a forthcoming publication \cite{todo}.

\end{document}